\documentclass[aps,prl,twocolumn]{revtex4-1}
\usepackage{amsmath}
\usepackage[usenames]{color}
\usepackage{float}
\usepackage{graphicx}
\usepackage{hyperref}
\usepackage{wrapfig}
\usepackage{amssymb}
\usepackage{enumitem}
\usepackage{bm}
\usepackage{subfigure}
\usepackage{placeins}
\usepackage{gensymb}

\begin{document}

\title{A DMRG study: Pair-density wave in spin-valley locked systems}

\author{Jordan Venderley and Eun-Ah Kim}
\affiliation{Cornell University, Ithaca, NY  14850}
\date{\today}

\begin{abstract}
Interest in modulated paired states, long sought since the first proposals by \textcite{Fulde} and by \textcite{Larkin}, has grown recently in the context of strongly coupled superconductors under the name of pair density wave (PDW). However, there is little theoretical understanding of how such a state might arise out of strong coupling physics in simple models. Although density matrix renormalization group (DRMG) has been a powerful tool for exploring strong coupling modulation phenomena of spin and charge stripe in the Hubbard model and the t-J model, there has been no numerical evidence of PDW within these models using DMRG.
Here we note that a system with inversion breaking, C$_{3v}$ point group symmetry may host a PDW-like state. Motivated by the fact that spin-valley locked band structure of hole-doped group VI transition metal dichalcogenides (TMD's) materializes such a setting, we use DMRG to study the superconducting tendencies in spin-valley locked systems with strong short-ranged repulsion. Remarkably we find robust evidence for a PDW and the first of such evidence within DMRG studies of a simple fermionic model.   

\end{abstract}

\maketitle

Recent experimental and theoretical developments have brought a renaissance to the idea of a modulated superconducting state that spontaneously breaks translational symmetry (see Ref.~\cite{Tranquada2015} and references therein and Refs.~\cite{PhysRevLett.99.127003,IntertwiningOrder,Seamus,cooper2017,PAlee2014}). 
Earlier efforts towards realization of modulated superonductors\cite{FFLOspinImbaColdgas,3DLOZeemanSOCgas} or towards an interpretation of associated experiments\cite{FFLOrashbascAgterberg} have  relied on generating finite-momentum pairing using spin-imbalance under an (effective) magnetic field, 
in close keeping with the original proposals\cite{Fulde,Larkin} (FFLO).
Alternatively, momentum space split, spinless fermions in the context of doped Weyl semi-metals\cite{WeylFFLOCho} and hole-doped transition metal dichalcogenides\cite{yiting} have been proposed as a platform for modulated superconductors due to pairing within fermi pockets centered at finite crystal momentum. 
On the other hand, a modulated paired state proposed for cuprates requires a strong coupling mechanism beyond Fermi surface effects.\cite{PhysRevLett.99.127003} Such a strong coupling driven state has been dubbed a PDW as a state distinct from FFLO-type superconductors. 

The need for a strong coupling mechanism without Fermi surface effects led to a search for the PDW state in numerical simulations. 
Numerous variational and mean-field studies have shown that pair-density wave type states are energetically competitive with uniform d-wave superconducting states in generalized t-J models, and it is thought that PDWs may become favorable in the presence of anisotropy.\cite{Lodor2011,Ogata2009,Raczkowski2007,Yang2009,Poilblanc2010}
Nevertheless, numerical evidence from the controlled approach of DMRG is lacking within simple fermionic models as the only evidence of PDW within DMRG was established in the 1D Kondo-Heisenberg model\cite{BergPDW}. One signature difficulty in such a realization is that DMRG calculations on a Hubbard or t-J model on a square lattice with spin-rotation symmetry often find spin and charge stripe ground states instead of the PDW state. 
However, one could hope that frustrating spin order might nudge systems into a PDW state. Here we turn to a Hubbard model on the frustrated triangular lattice with broken inversion symmetry that captures the hole-doped monolayer group IV transition metal dichalcogenides (TMD's). 

Rapidly growing interest in the monolayer group VI transition metal dichalcogenides (TMD's) has been fueled by the exotic possibilities driven by spin-orbit coupling and lack of centrosymmetry\cite{Exp_IsingscMoS2_Iwasa,Exp_IsingscMoS2_LawYe,IsingNbSe2,yiting,Bawden,TMD2012,SpinHall} as well as superconductivity in the n-doped TMD's\cite{ndoped1,ndoped2,ndoped3,Exp_IsingscMoS2_Iwasa,Exp_IsingscMoS2_LawYe,IsingNbSe2}.
While the symmetry properties of the observed superconducting states remain unknown, the different translationally invariant superconducting channels for the TMD's have been previously classified in mean-field studies \cite{KTLaw,pdopedTMDAji,Guinea}. 
Recently \textcite{yiting} employed a weak-coupling RG approach 
to investigate a repulsive interaction driven pairing mechanism, predicting 
two topological superconducting instabilities with one of them being a spatially modulated intra-pocket state. However potentially strong correlation effects have largely been neglected despite the fact that the conduction electrons have substantial \textit{d}-character.
In this letter we use density matrix renormalization group (DMRG) calculations to study the effects of spin-orbit coupling on superconducting tendencies driven by repulsive interactions. 

 
DMRG is a powerful, non-perturbative method for studying strongly interacting systems \cite{DMRGinAgeofMPS,KagomeHeisenberg,Stripes,StripesPairing,BergPDW,Dplusid}. It has been used with great success to explore a diverse selection of strongly correlated phenomena highlighted by stripes, spin-liquids, and superconductivity \cite{KagomeHeisenberg,Stripes,StripesPairing,BergPDW,Dplusid,SpinLiquid,ChiralSpinLiquid,DwaveMetal,EntanglementEntropyDMRG,J1J2}. However, since DMRG is quasi-1D in nature no true long-range order can be seen in the correlations. Thus in order to access our system's superconducting tendencies we implement a pair-edge-field motivated by the field-pinning approach underlying several earlier studies \cite{stripedttj,2DimDMRG,KagomeHeisenberg,BergPDW}. By biasing the system towards a particular superconducing state and studying the emergent symmetry of the appropriate order parameter in the bulk, one can gauge the model's propensity for various instabilities. 


\begin{figure}[t]
\subfigure[]{
	\raisebox{+0.6\height}{\includegraphics[width=0.45\linewidth]{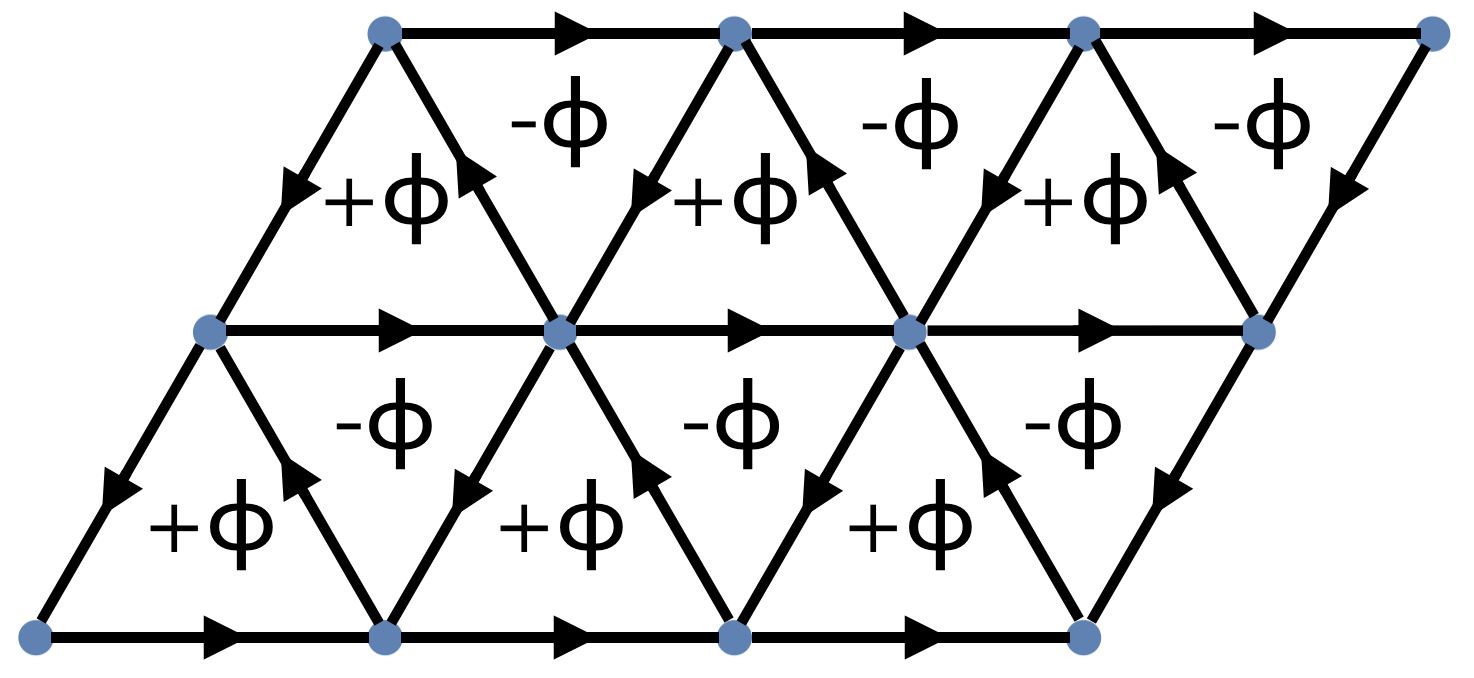}}
	\label{fig:fluxlattice}}
\subfigure[]{
	\includegraphics[width=0.48\linewidth]{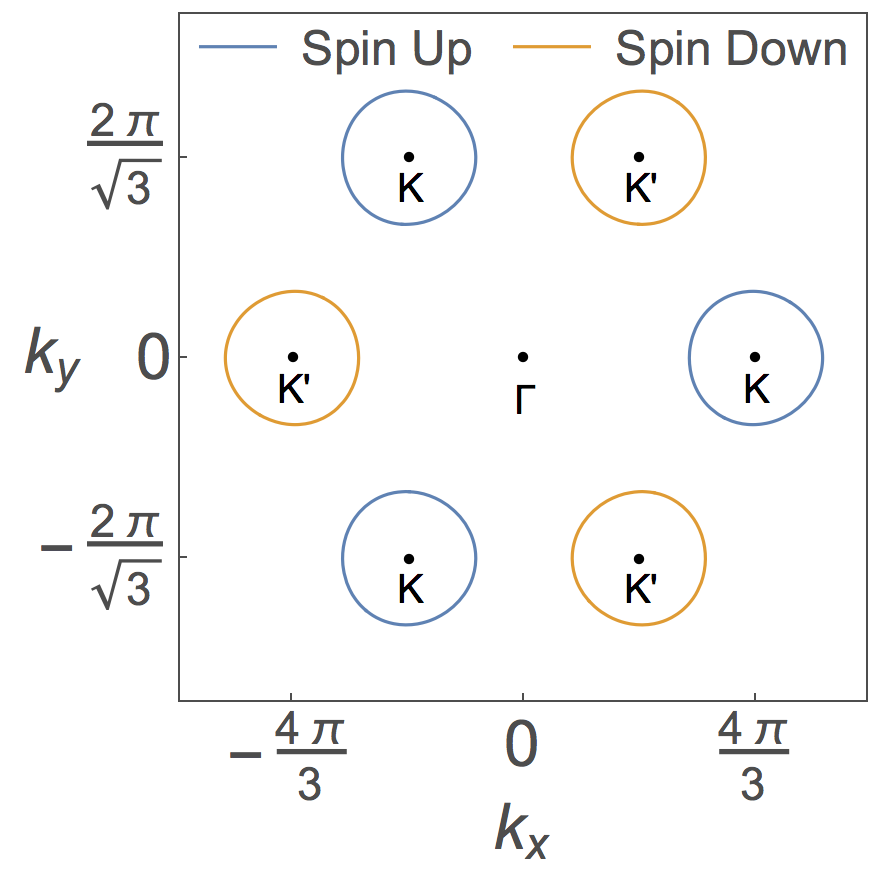}
	\label{fig:pockets}}
\caption{(a) The spin-dependent staggered flux pattern for one spin component with $\pm \Phi$ flux per plaquette. An opposite flux pattern for the other spin component guarantees time-reversal symmetry The arrows indicate the direction of positive phase hopping. (b) Our Fermi surface with \mbox{$t_{i+\hat{x}, i;\uparrow} = \frac{2}{\sqrt{3}}e^{i 0.3 \pi }$} and $\mu =4.6$ in the tight-binding model in Eqs: \eqref{eq:H} \eqref{eq:dispersion}. Here the spin-valley locked, circular Fermi pockets are evident.}
\label{bandcomp}
\end{figure}

In order to capture the spin-valley locked Fermi surfaces that occur in the valence band of group VI TMD's \cite{Exp_IsingscMoS2_Iwasa,mos2abinitio,5band} in a one-band model we consider a 
nearest-neighbor tight-binding model on a triangular lattice with a staggered, spin-dependent magnetic flux of $\pm 0.9 \pi$ per plaquette (see Figure~\ref{fig:fluxlattice}).
This spin-dependent flux breaks the 
C$_{6v}$ symmetry down to C$_{3v}$ while preserving the time-reversal symmetry, mimicking the $S_z$ preserving spin-orbit coupling present in the Kane-Mele model and generating two distinct spin-polarized pockets for our Fermi surface. Furthermore, the flux introduces a small amount of anisotropy in the pockets (see Fig~\ref{fig:pockets}) analogous to that present in real materials such as MoS$_2$.\cite{Kormanyos2014}
Finally, we include an on-site interaction $U$. Hence our model Hamiltonian is 
\begin{equation}
H= -\sum\limits_{\langle i,j\rangle} t_{ij,\sigma}c^{\dagger}_{i\sigma} c_{j\sigma} -\mu \sum\limits_{i\sigma} c^{\dagger}_{i\sigma} c_{i\sigma} + U \sum\limits_i n_{i\uparrow} n_{i\downarrow},
\label{eq:H}
\end{equation}
where $t_{ij,\sigma}$ is the spin-dependent complex nearest-neighbor hopping 
, $\mu$ is the chemical potential, and U is an on-site Hubbard interaction.  
The dispersion takes the form of 
\begin{equation}
\epsilon_{\sigma}(\bm{k}) = -2\sum\limits_i \Big[ \text{Re}(t) \cos(\bm{\delta}_i \cdot \bm{k}) + \text{Im}(t) \sin(\bm{\delta}_i \cdot \bm{k}) \sigma_z \Big]
\label{eq:dispersion}
\end{equation}
where $\delta_i \in \{ \hat{\mathbf{x}}, -\frac{1}{2}\hat{\mathbf{x}} +\frac{\sqrt{3}}{2}\hat{\mathbf{y}},  -\frac{1}{2}\hat{\mathbf{x}} -\frac{\sqrt{3}}{2}\hat{\mathbf{y}} \}$, $\sigma_z = \pm$ for spin up and down respectively, we define $t$ such that $t = t_{i+\hat{x}, i;\uparrow}$ , and the lattice spacing has been set to 1. 

Even in the absence of the spin-valley locking special to our model, the geometric frustration of the triangular lattice is known to foster exotic phases.
While a consensus has emerged that the ground state of the Heisenberg model is a 120\degree N\'eel antiferromagnet\cite{White2007}, the Hubbard model has been shown to have tendencies toward spin liquid and chiral d+id superconducting states.\cite{Sahebsara2008,Chen2013}. It is notable that within the context of the Heisenberg model frustration can inhibit spin-stripes.

We emphasize that due to the lack of inversion symmetry, even and odd pairing components can coexist\cite{RMPUnconvScSigrist}. Thus the $S_z$ preserving spin-orbit coupling allows for the mixing of $S_z=0$ singlet and triplet states, i.e., the bond pair order parameter $\Delta_{ij} = \langle c^{\dagger}_{i\uparrow}c^{\dagger}_{j\downarrow} \rangle$ should be an admixture of singlet and triplet components: 
\begin{align}
\begin{split}
\Delta_{ij}^{singlet} &= \langle c^{\dagger}_{i\uparrow}c^{\dagger}_{j\downarrow} - c^{\dagger}_{i\downarrow}c^{\dagger}_{j\uparrow} \rangle \\
\Delta_{ij}^{triplet} &= \langle c^{\dagger}_{i\uparrow}c^{\dagger}_{j\downarrow} + c^{\dagger}_{i\downarrow}c^{\dagger}_{j\uparrow} \rangle
\label{eq:orderparam}
\end{split}
\end{align}
Note that since our system lacks translational symmetry due to open boundary conditions, the pairing symmetry is not constrained to transform under a single irreducible representation. Nevertheless, the real-space structure of these bond-centered order parameters provides insight into the nature of the dominant pairing state.

\begin{figure}[t]
\begin{center}
\includegraphics[width=\linewidth]{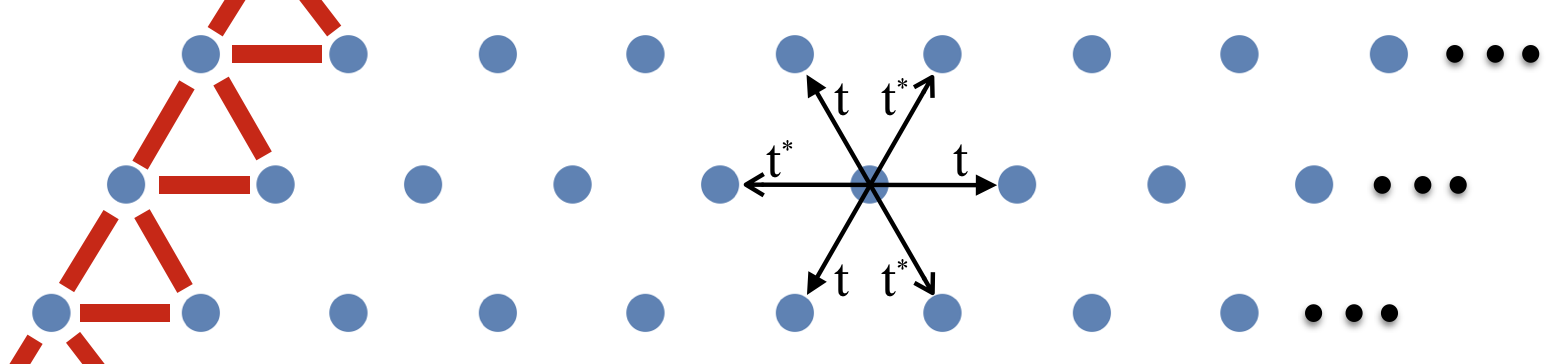}
\caption{\label{fig:lattice} A depiction of our lattice. It is periodic in the short direction with 3 unit cells and has open boundaries in the long direction. The ellipses on the right signify that multiple lengths are studied: L = 12, 18, 24, 36. The edge field, shown as red lines, is a pair-field of the form given in equation \ref{eqn:edge-field}. The nearest-neighbor hopping structure for spin up is also shown with the spin down hopping structure being the complex conjugate of that shown above.}
\end{center}
\end{figure}

We carry out our DMRG simulations on a cylinder with 3 unit cells in the periodic direction and 12, 18, 24, and 36 unit cells in the non-periodic direction. The width is sufficiently large to sample both types of pockets in the Fermi surface but not so large as to make DMRG prohibitively expensive for our available computational resources. We keep the band structure fixed and explore the effects of varying $U$.
We investigate the superconducting susceptibility by applying a pair-field along one edge as illustrated above in Fig.~\ref{fig:lattice}.
In order to reveal any inherent preferences for a particular superconducting channel, we consider two different phase structures for the edge field: a uniform field described by the A1 irrep and a random field i.e. 
\begin{equation}
\Delta^{edge}_{ij}= V c^{\dagger}_{i\uparrow}c^{\dagger}_{j\downarrow} e^{i \phi_{ij}} + \text{h.c.}
\label{eqn:edge-field}
\end{equation}
with the phase $\phi_{ij}$ chosen to transform under the A1 irrep or be random for $i \ne j$ and $0$ for $i=j$. We remark that all results presented in this paper have been shown to be independent of the phase structure of the edge field applied [see Appendix I]. The strength of the pair-field was fixed to be V=$0.1$, about an order of magnitude less than the hopping amplitude which is consistent with that used in previous studies.\cite{stripedttj,2DimDMRG,BergPDW}
While the cylindricity of our geometry and the addition of a pair field break the C$_{3v}$ symmetry of the lattice and translational invariance, if there is a well-defined structure to order parameter in the bulk, we expect to gain information about the inherent superconducting tendencies through the real space structure of the order parameters in Eq.~\eqref{eq:orderparam}.

Our DMRG calculation utilizes the iTensor library developed by Miles Stoudenmire and Steve White.\cite{itensor} We perform up to 14 sweeps with a final bond dimension of M=2500. This is sufficient to obtain energy convergence to $\mathcal{O}(10^{-8})$ for our repulsive calculations and  $\mathcal{O}(10^{-10})$ for attractive calculations [see Appendix II]. We focus exclusively on the inter-pocket instabilities of our model where we may exploit the conservation of the S$_z$=0 quantum number. S$_z$ and fermion parity, N$_2$, are conserved quantum numbers, but the U(1) particle number symmetry is broken by the pair field.  As our starting point, we construct a MPS that randomly samples the S$_z$=N$_2$=0 sector of our Hilbert space picking 30 states from each even particle number sector. In addition, we use exact diagonalization (ED) to ensure correctness in the DMRG simulations with the energetics from the two methods agreeing to within machine precision for small 3x3 systems where ED is computationally tractable.

\begin{figure}[H]
\centering
\subfigure[]{
    \includegraphics[width=\linewidth]{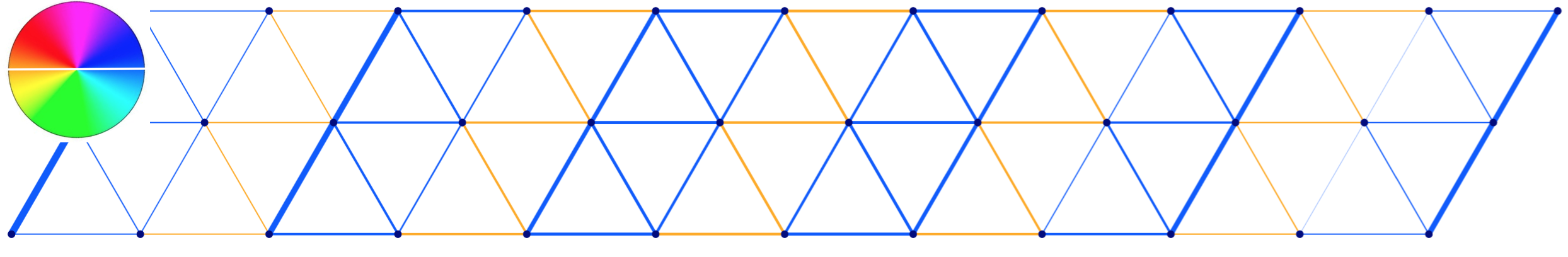}
    \label{fig:L24cut_U2_mu4p6_phase_singlet}}
\subfigure[]{
\includegraphics[width=0.85\linewidth]{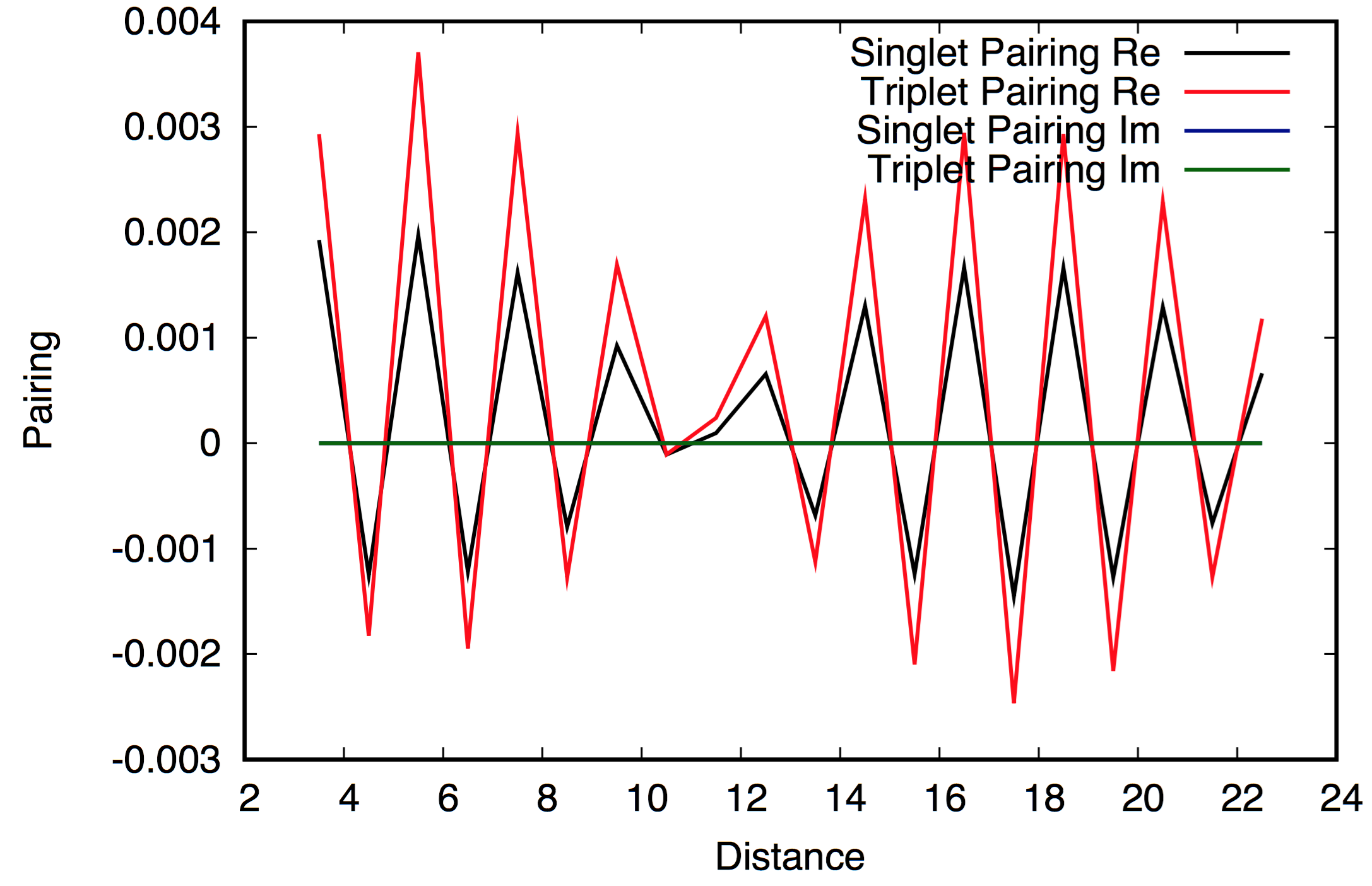}
    \label{fig:sing_trip_lin}}
\caption{(a) Arg$(\Delta^{singlet}_{\langle i,j \rangle})$ for all nearest-neighbors with $U = +2$ for our $3\times 24$ lattice with periodic boundary conditions along the short direction and open boundary conditions along the long direction. For visibility, we truncate the plot so that only the half farthest from the edge field is shown. The line thickness is proportional to the pairing amplitude. (b) We plot the real and imaginary components of $\Delta^{singlet}_{ij}$ and $\Delta^{triplet}_{ij}$ for i,j along the middle rung of our lattice in order to present the phase oscillations }
\label{fig:repulsive-pair}
\end{figure}

We initially consider a moderately attractive interaction with U=$-$2, where uniform pairing in the A1-irrep is expected. Data for the following statements may be found in Appendix III. 
First, we note that the phase disorder due to edge field effects
quickly disappears upon moving into the bulk [see Fig. A\ref{fig:L36cutleft_Uneg2_mu4p6_phase_singlet} in Appendix I]. Inspecting the bond-singlet component of the superconducting order parameter $\Delta^{singlet}_{\langle ij \rangle}$ on the directed nearest-neighbor bonds away from the probe, 
we find a well-ordered uniform phase structure, Fig. A\ref{fig:L36cut_Uneg2_mu4p6_phase_singlet}. 
Moreover, zooming into the phase structure within a single unit-cell, we find the pair-field expectation value to be isotropic and definitively s-wave.
The uniform and isotropic nature of the order parameter phase distribution is a robust property of our results in the negative $U$ regime, that is insensitive to the profile of the edge fields and occurs for all system sizes studied. The triplet channel behaves analogously, displaying homogeneous f-wave pairing [see Fig. A\ref{fig:L36cut_Uneg2_mu4p6_phase_triplet} in Appendix III].

Armed with the attractive, $U<0$, result that can serve as a reference, we now study the moderately repulsive Hubbard regime with $U=+2$. Earlier work using two-stage perturbative RG on a similar spin-valley locked model with repulsive $U$ predicted superconductivity in the two dimensional $E$ representation where some linear combination of \textit{p} and \textit{d}-wave symmetries occurs due to the lack of inversion symmetry\cite{yiting}. Unexpectedly, our DMRG simulation in this repulsive interaction regime 
reveals a tendency to break translational symmetry along the length of the cylinder. Specifically the system forms a modulated paired state where both the singlet and triplet bond pair order parameters are everywhere real with modulation in their sign (see Fig.~\ref{fig:repulsive-pair}). From the symmetry perspective the observed state is analogous to the state proposed by \cite{Larkin}. 
Such modulation in the pair amplitude is evident in  the plot of $\Delta^{singlet}_{\langle ij \rangle}$ for $U = +2$ in Fig.~\ref{fig:repulsive-pair}(a) where an anisotropic phase structure within the unit-cell is repeated with period 2. A similar unit-cell doubling is seen in the triplet channel [see Appendix IV]. We find this tendency to form a PDW is robust against changes in chemical potential although the periodicity depends on the chemical potential in a non-trivial manner [see Appendix V]. For instance, increasing the 
chemical potential, $\mu$, from $\mu = 4.6$ to $\mu = 6.0$ enlarges the unit-cell by an additional lattice site [see Fig. \ref{fig:L36cut_U2_mu6_phase_singlet}].
Although there has been much interest in modulated superconducting states, 
this is the first report of a strong coupling driven PDW within DMRG simulations to the best of our knowledge. 

To gain further insight into the observed PDW phenomena, we compare oscillations in the singlet pairing strength and in the bond charge density of the attractive case to those in the repulsive one. Since $\Delta^{singlet}_{ij}$ is characterized exclusively by $\pi$-phase shifts in the bulk, we may project it to real space after a gauge transformation and look at the decay properties of the pair field by plotting it for bonds directed along the middle rung, see Fig. \ref{fig:repulsive-pair}(b). For attractive interactions, the singlet pairing strength falls off gradually as expected from the quasi-1D geometry of the system and exhibits slowly varing oscillations [see Appendix III Fig. \ref{fig:sing_trip_lin_att}] due to finite size effects induced by the open boundary conditions\cite{Friedel}. On the other hand, the pairing profile for the $U=+2$ simulation shows additional rapid oscillation with period 2, see Fig. \ref{fig:repulsive-pair}(b). While the number of enveloping ``beats" in the pairing profile is dependent on system size, these oscillations about zero occur for all repulsive Hubbard simulations near $U=+2$. Thus although the exact strength of the pairing response and the penentration depth of the edge field appear to have some dependence on the edge field profile and the length of the lattice, the PDW-type behavior reported has been observed for all system sizes and all edge-field types. Intriguingly this plot strongly resembles the plot of the same quantity in the Kondo-Heisenberg model with PDW\cite{BergPDW}.

\begin{figure}[ht]
\centering
    \includegraphics[width=0.75\linewidth]{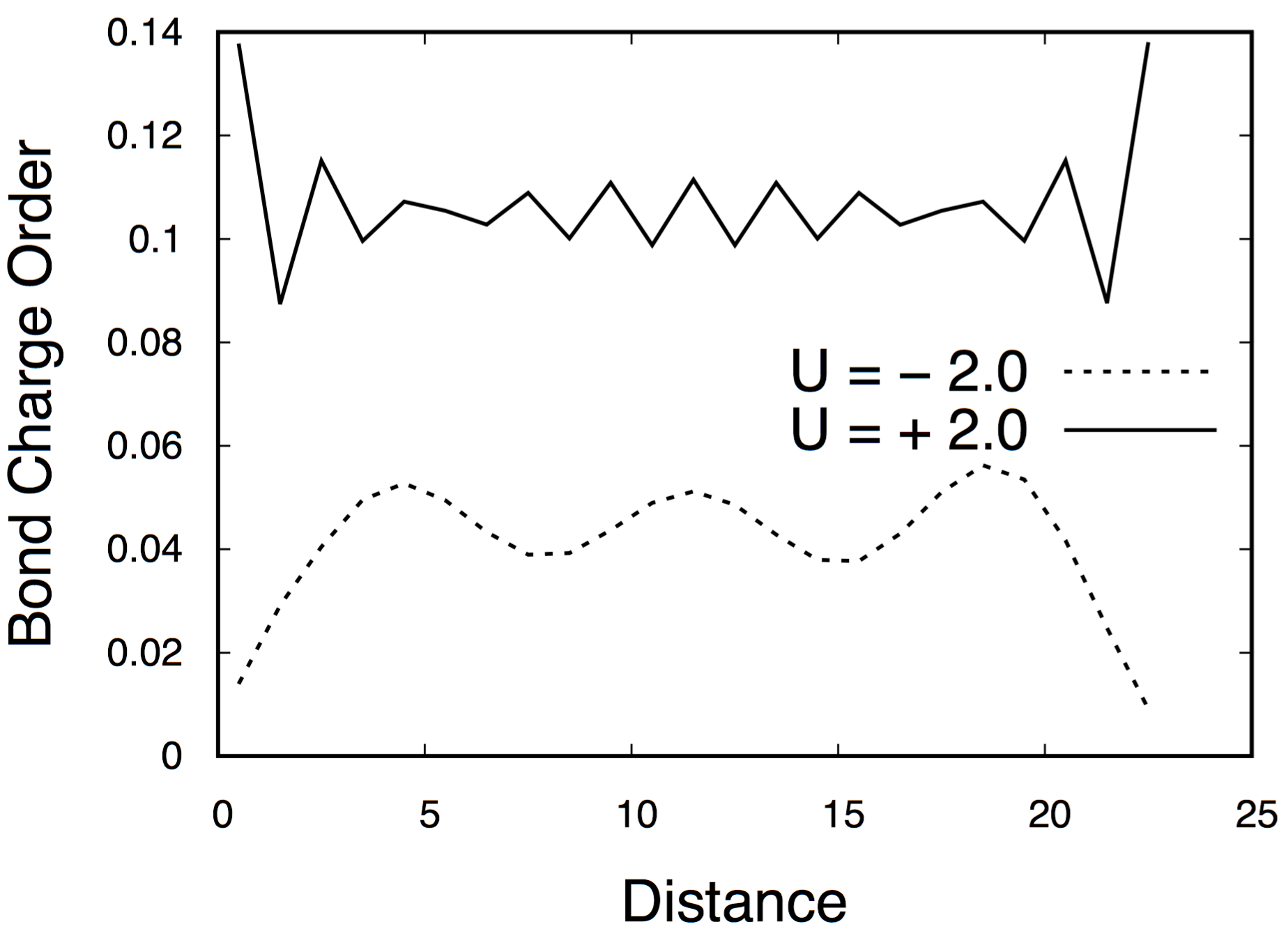}
\caption{Bond charge order, $\sum\limits_{\sigma}\langle c^{\dagger}_{i\sigma} c_{j\sigma} + c^{\dagger}_{j\sigma} c_{i\sigma} \rangle$ for nearest-neighbor sites, $i$ and $j$, lying along the middle rung. The bond-center coordinate is used for the x-axis. }
\label{charge_bond}
\end{figure}

As the pairing amplitude profile of our $U=+2$ simulation has net pair amplitude on the whole system 
due to the edge field, charge modulation of the same period (period 2) is anticipated to be driven by the net component and the modulated pairing components\cite{Tranquada2015}. Indeed the bond charge density profile for $U=+2$ shows the anticipated rapid oscillation on top of the slowly varying standing wave profile [see Fig. \ref{charge_bond}]. We remark that while both the attractive and repulsive charge bond densities have oscillations that dip below their mean values only the superconducting response of the repulsive case has oscillations about zero suggesting that the PDW-like phase shifts are not finite size effects.

In summary, we used DMRG to study the superconducting tendencies of a repulsive-$U$ Hubbard model on a triangular lattice with spin-valley locking. These tendencies were probed by studying the pairing response profile in response to uniform and 
random pair fields along one edge.  
Our calculations indicate that the superconducting phase diagram of the model may be more complex than what was revealed by the previous perturbative RG study \cite{yiting}, with translational symmetry breaking superconducting states possibly in competition with a uniform state. The PDW observed 
breaks translational symmetry with the superconducting order parameter alternating sign. 
This atypical pairing response may be related to the fact that Ising spin-orbit coupling and triangular lattice consipire to frustrate any spin order including spin stripe. This conjecture is supported by the fact that we observe $S_z$-$S_z$ correlations to be predominantly negative without any sign of stripe response, a feature characteristic of frustrated spin systems.  
Moreover, the fact that the apparent periodicity is not related to any Fermi surface properties suggest a purely strong coupling origin of the observed translational symmetry breaking. 
It will be interesting to study whether the observed PDW state can be found in a purely two-dimensional setting using a different method.

{\bf Acknowledgements} We thank Leon Balents, Erez Berg, Garnet Chan, Yi-Ting Hsu, Steve Kivelson, Kam Tuan Law, Patrick Lee, Steve White, and  Hong Yao 
for useful discussions. E-AK acknowledges Simons Fellow in Theoretical Physics Award $\#$392182 and DOE support under Award de-sc0010313. E-AK is grateful to the hospitality of Kavali Institute of Theoretical Physics supported by NSF under Grant No. NSF PHY-1125915, where this work was completed. 
JV acknowledges support by the National Science Foundation (Platform for the Accelerated Realization, Analysis, and Discovery of Interface Materials (PARADIM)) under Cooperative Agreement No. DMR-1539918 and in part by NSF DMR-1308089.



\bibliographystyle{apsrev4-1}
\bibliography{sources}

\clearpage
\appendix
\setcounter{figure}{0} \renewcommand{\thefigure}{A.\arabic{figure}}

\begin{widetext}

\section{Appendix I: Phase Structure Near the Random Pair-Field}

Since in the main text we only provide the phase structure of the superconducting order parameter farthest from the pair-field, we now show the edge closest to the probe for the arguably more interesting case where the edge field has a random phase structure . We only provide the singlet plots as evidence but the triplet channel behaves analogously. These plots illustrate the effect of the random pair-field in inducing randomness into the pairing phase and highlight the tendency for the pairing to settle into a dominant phase far from the edge field.

\begin{figure}[H]
\centering
\subfigure[]{
   \hspace{40pt} \includegraphics[width=0.75\linewidth]{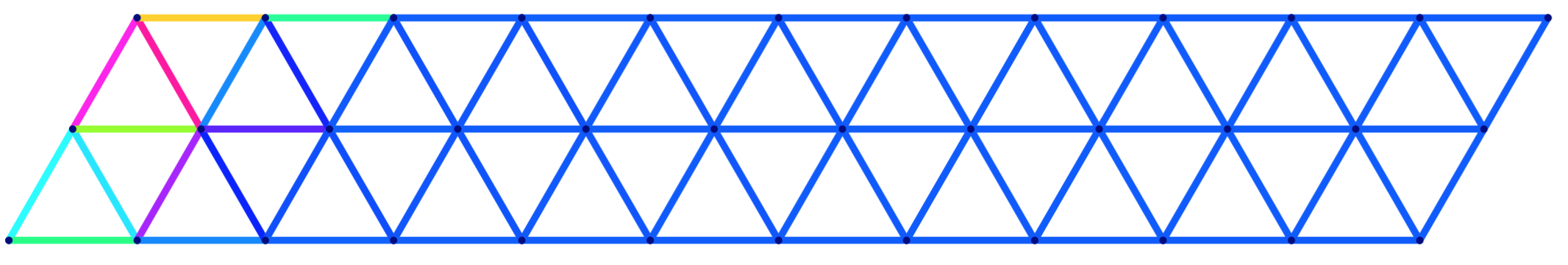}
    \label{fig:L36cutleft_Uneg2_mu4p6_phase_singlet}}
 \subfigure[]{
   \hspace{40pt} \includegraphics[width=0.75\linewidth]{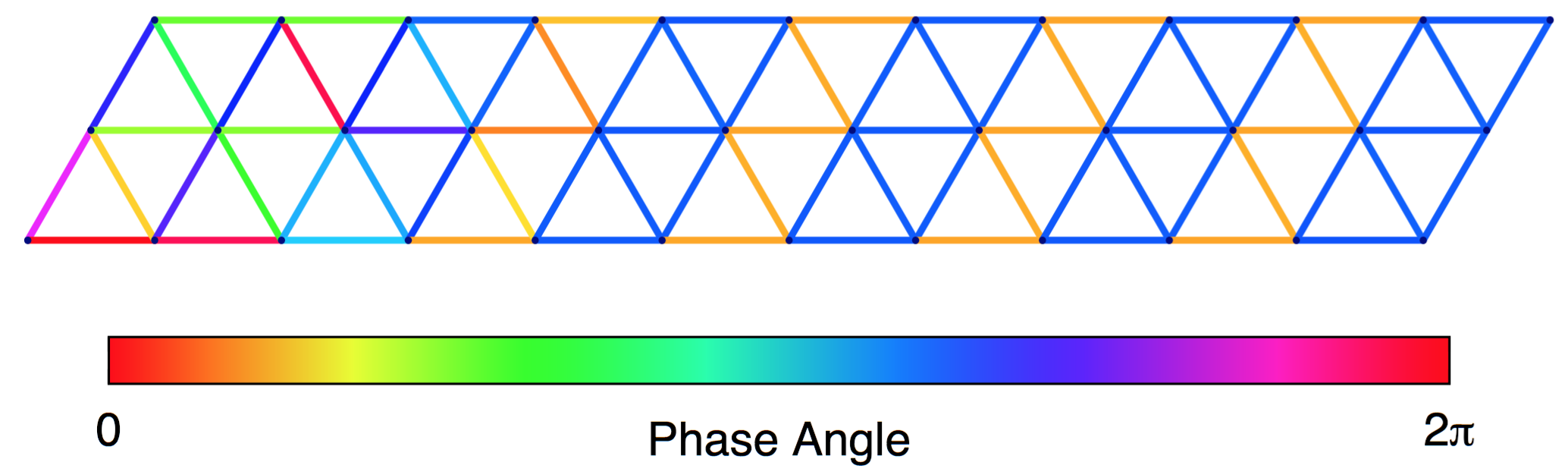}
    \label{fig:L36cutleft_U2_mu4p6_phase_singlet}}
\begin{picture}(0,0)
\put(-480,170){\huge{U = $-$2}}
\put(-480,70){\huge{U = +2}}
\end{picture}
\caption{The singlet phase structure, Arg$(\Delta^{singlet}_{\langle i,j \rangle})$, for (a) the attractive, $U = -2$, and (b) the repulsive, $U = +2$, for the L = 36 lattice with random pair-edge-field where now we plot only the third closest to the pair-field. As before, the lattice has periodic boundary conditions along the short direction and open boundary conditions along the long direction. Unlike previous plots of this kind, the line thickness here is constant since we only want to emphasize the phase change and the decay from the edge makes this difficult to see.}
\label{fig:left-pairing-phases}
\end{figure}
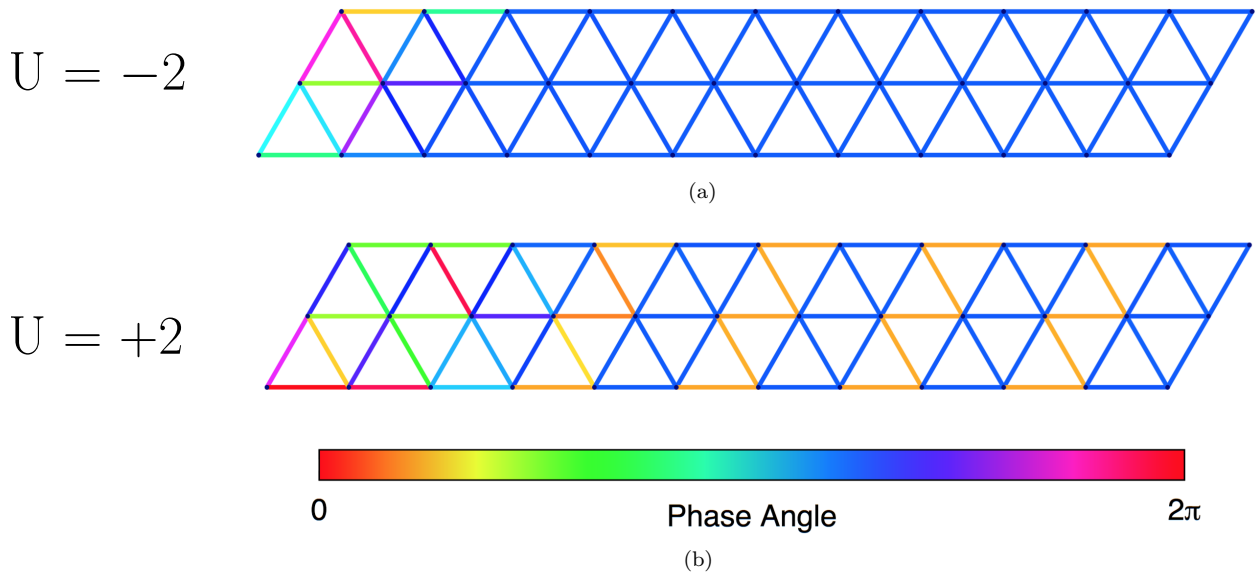

\section{Appendix II: DMRG convergence}

We check the convergence of our DMRG simulations by looking at the change in energy between sweeps. For all simulations we start our first sweep with a maximum bond dimension of 500 states and by the 14th sweep keep up to 2500 states, the maximum allowed by our current RAM limitations. We specify an SVD truncation error of $10^{-12}$ but in practice find that convergence within iTensor is limited to $\mathcal{O}(10^{-10})$. Typical convergence is achieved to $\mathcal{O}(10^{-8})$ for repulsive interactions and $\mathcal{O}(10^{-10})$ for attractive interactions.  Below we plot the decrease in energy between sweeps for the L = 24, U = +2 calculation presented in the main text.

\begin{figure}[H]
\begin{center}
\includegraphics[width=.3\linewidth]{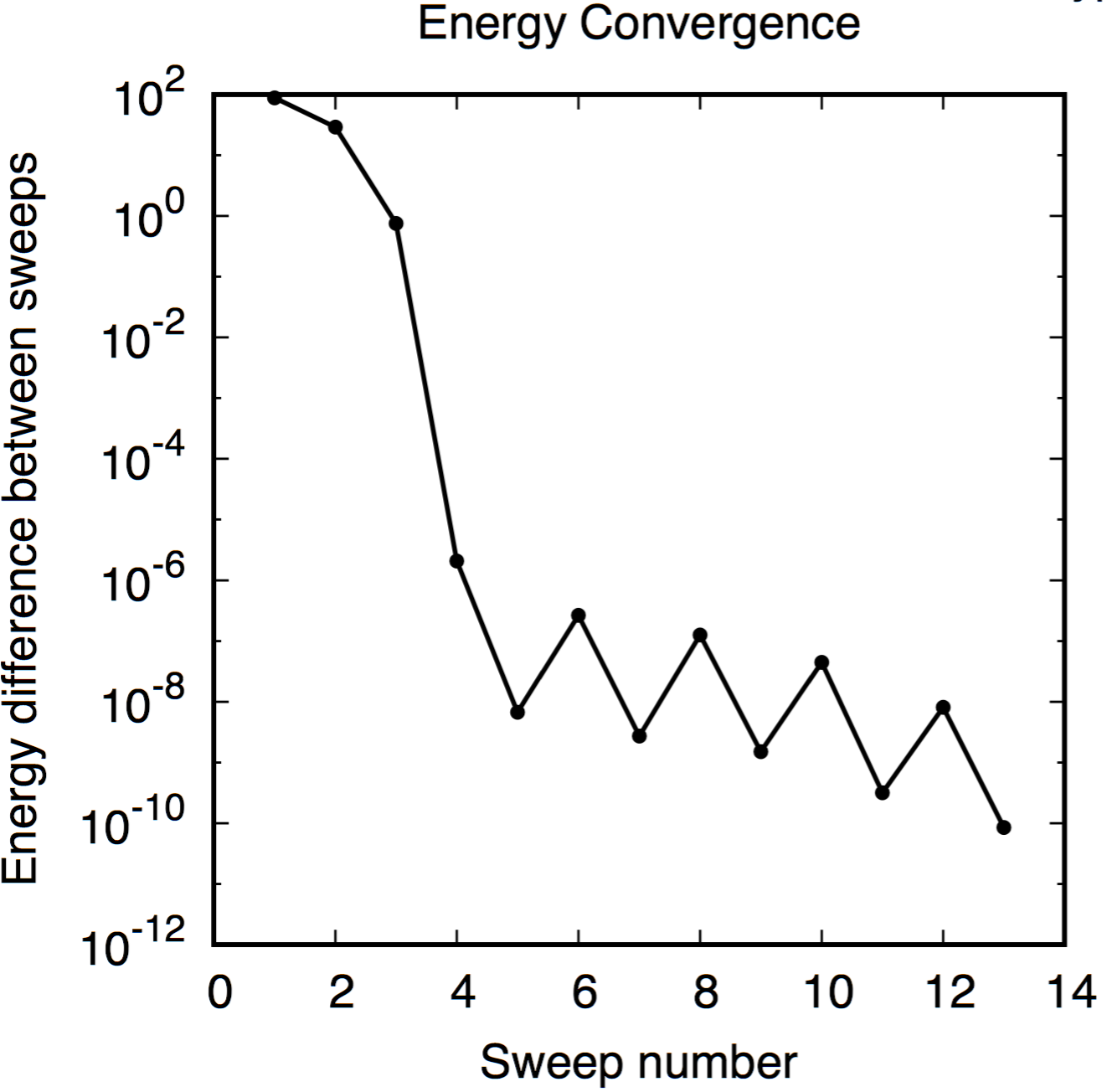}
\caption{\label{fig:lattice} The decrease in energy between sweeps for the DMRG caclulation in the main text (L = 24, U = +2) demonstrating convergence.}
\end{center}
\end{figure}

\section{Appendix III: Phase Structure for Attractive Interactions} In order to establish a reference for our pair-edge field approach, we explore the effect of an attractive Hubbard interaction, U = -2, on the superconducting preferences. More specifically, we look at the bond-singlet and bond-triplet components of the superconducting order parameter along directed nearest-neighbor bonds. Here we find that regardless of the phase structure of the edge field, uniform or random, homogenous pairing is established in the bulk and any phase disorder due to the edge field
quickly disappears upon moving away from the edge field [see Fig. A\ref{fig:L36cutleft_Uneg2_mu4p6_phase_singlet} in Appendix I]. This induced, translationally invariant phase structure is found to be an admixture of \textit{s}-wave and \textit{f}-wave pairing, as expected for a conventional supeconductor breaking parity symmetry. The uniform-A1 behavior of the bulk order parameter is robust, being insensitive to the profile of the edge fields and system size. Note that we have opted to use a different scheme for presenting the phase here as compared to the main text since the triplet component cannot be presented using the previous approach. We use this alternative style whenever plotting triplet phases but include it also for the singlet channel here along with the conventional style in order to help understand what's being presented.

We also provide the pairing strength along the middle rung analogous to the plot in Fig. \ref{fig:sing_trip_lin}.Here we plot the real and imaginary components of $\Delta^{singlet}_{ij}$ and $\Delta^{triplet}_{ij}$ for attractive Hubbard interactions with i,j along the middle rung of our lattice. Comparison to the analogous plot in the main text for the repulsive case, Fig. \ref{fig:sing_trip_lin}, highlights the presence of oscillations about 0 in the repulsive case that are absent in the attractive case.

\begin{figure}[H]
\centering
\subfigure[]{
    \includegraphics[width=0.75\linewidth]{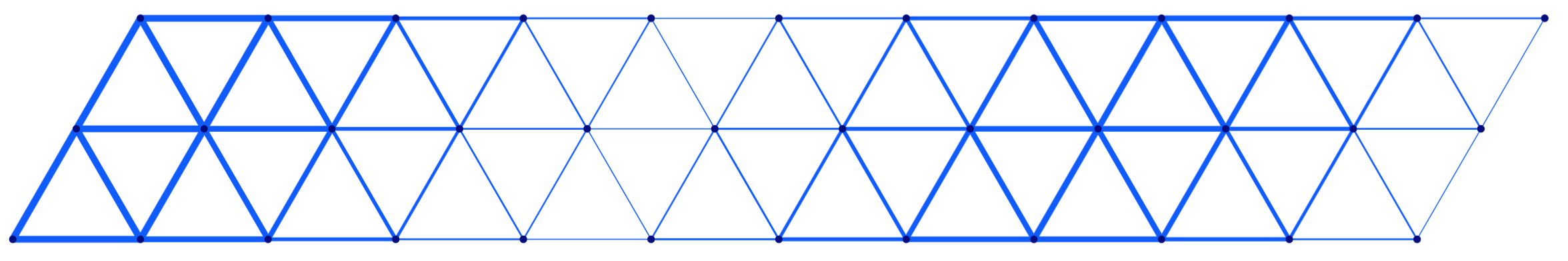}
    \label{fig:L36cut_Uneg2_mu4p6_phase_singlet}}
\subfigure[]{
    \includegraphics[width=0.75\linewidth]{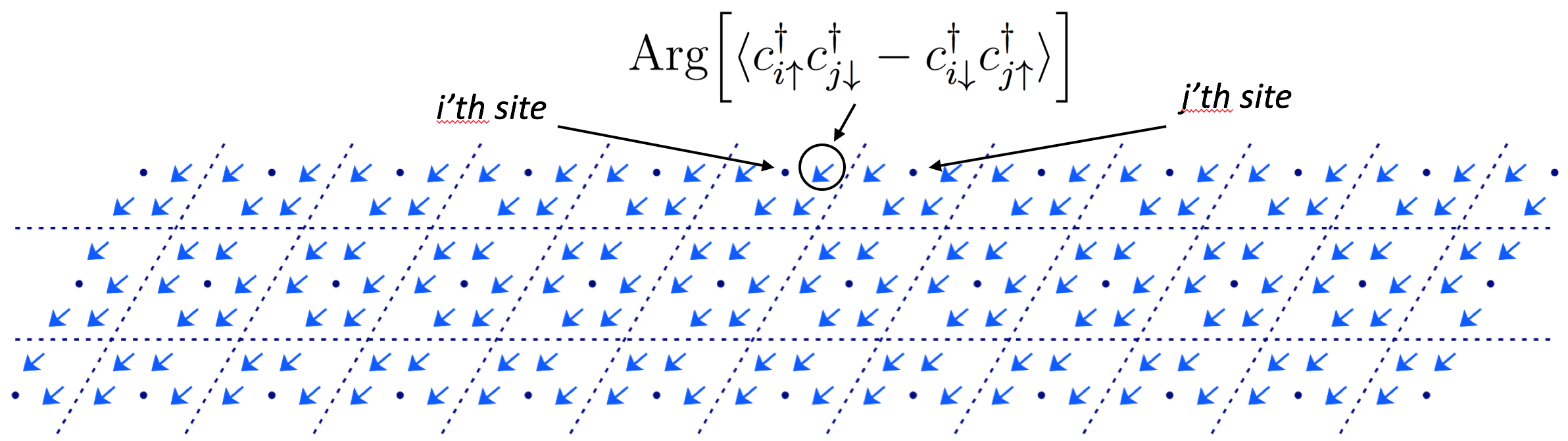}
    \label{fig:L36cut_Uneg2_mu4p6_phase_singlet}}
 \subfigure[]{
    \includegraphics[width=0.75\linewidth]{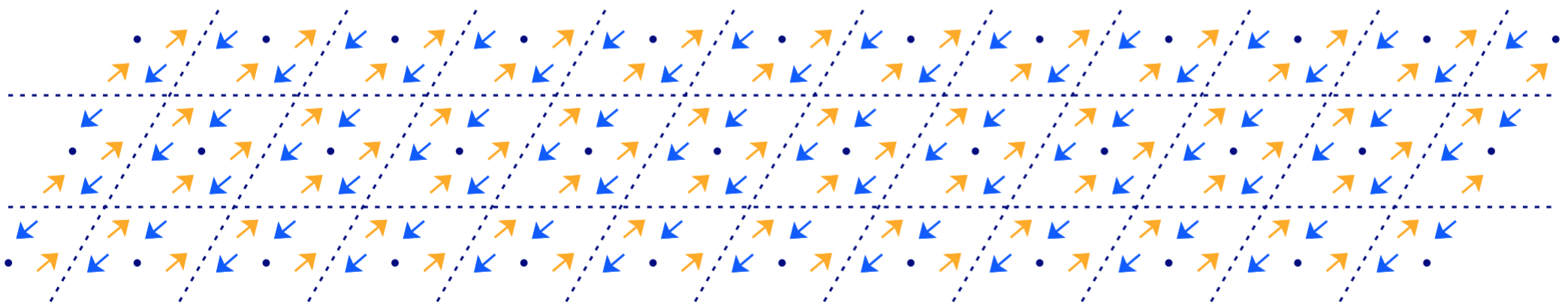}
    \label{fig:L36cut_Uneg2_mu4p6_phase_triplet}}
\subfigure{
    \includegraphics[width=0.85\linewidth]{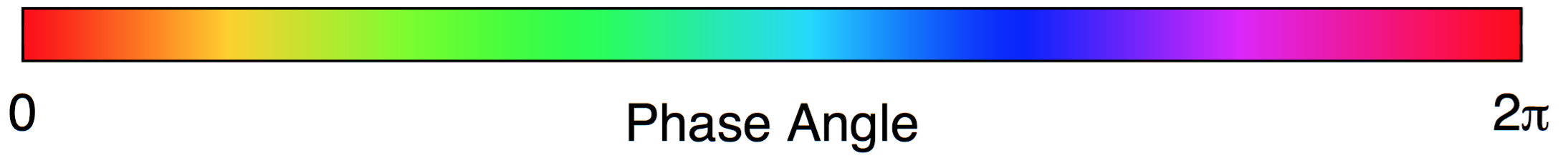}
    \label{fig:phase_legend}}
\caption{(a) $\Delta^{singlet}_{\langle ij \rangle})$, (b) Arg$(\Delta^{singlet}_{\langle ij \rangle})$ and (c) Arg$(\Delta^{triplet}_{\langle ij \rangle})$ for $U = -2$ along all directed nearest-neighbor bonds. Note that (a) and (b) presents the same phase data and (a) is given only to help the understanding of the new style of plotting. Here we provide the L=24 results, but only show the half of the lattice away from the edge probe. Recall our lattice has periodic boundary conditions along the short direction and open boundary conditions along the long direction. The arrows point in the direction of the phase and are also colored according to the argument. Although redundant, this is done to aid visibility. Note that this method of plotting possesses an additional redundancy in that $\Delta^{singlet}_{\langle ij \rangle} = \Delta^{singlet}_{\langle ji \rangle}$ and  $\Delta^{triplet}_{ij} = - \Delta^{triplet}_{ji}$.}
\label{fig:long-pairing-triplet-phases}
\end{figure}

\begin{figure}[H]
\centering
    \includegraphics[width=0.5\linewidth]{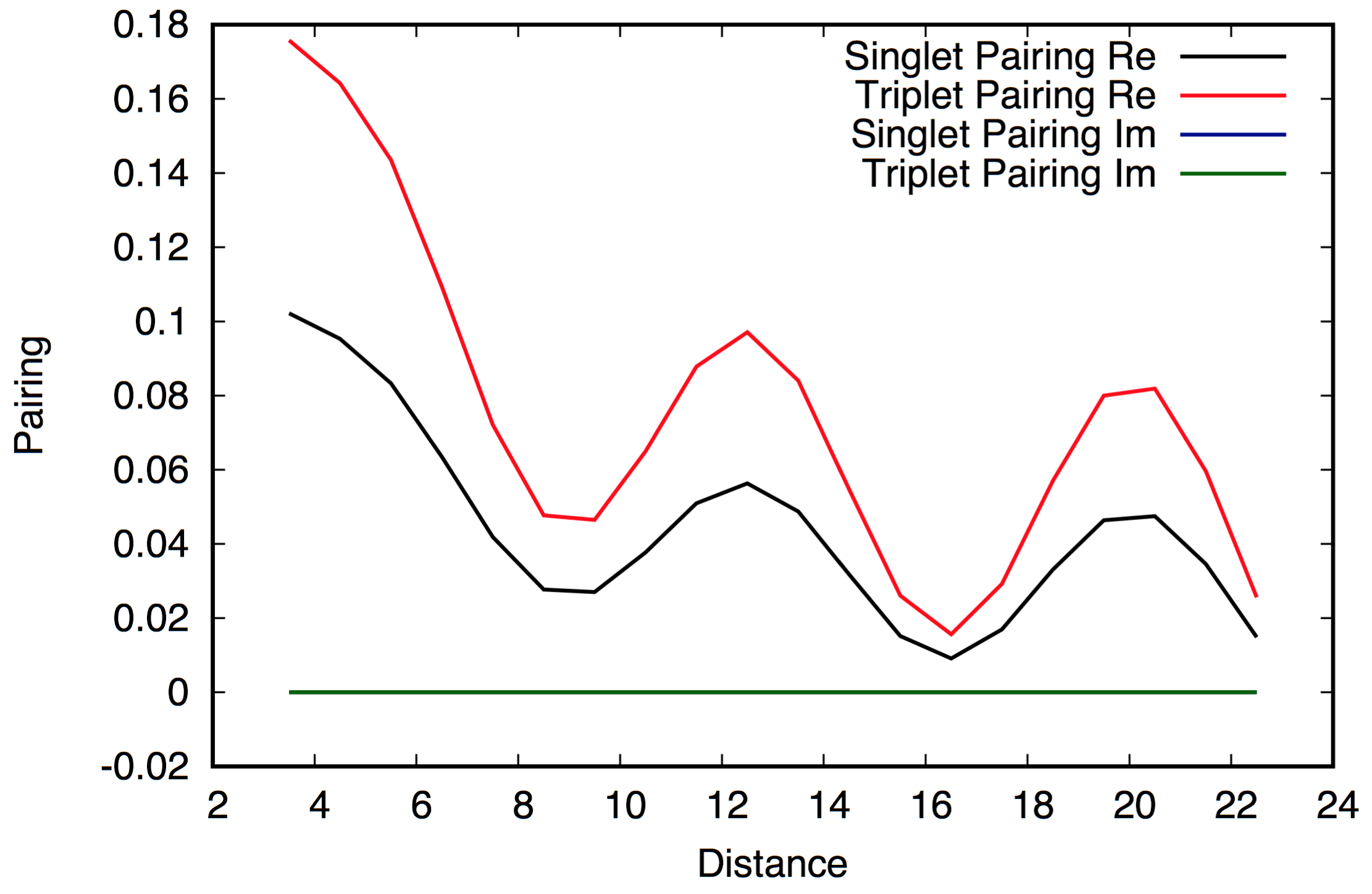}
\caption{The real and imaginary components of $\Delta^{singlet}_{ij}$ and $\Delta^{triplet}_{ij}$ for the attractive Hubbard regime, U = -2, of the L=24 lattice with uniform edge field. Here i,j lie along the middle rung of our lattice. }
\label{fig:sing_trip_lin_att}
\end{figure}

\section{Appendix IV: Triplet Phase Structure for Repulsive Interactions}
In Fig. \ref{fig:L36cut_U2_mu4p6_phase_triplet} we provide the triplet phase plot of the L=24 lattice with uniform edge field for U = +2 case shown in the main text. Again, for the sake of visibility, only the half of the lattice farthest from the edge field is displayed. This result is qualitatively similar to the corresponding singlet case shown in the main text, Fig.  \ref{fig:L24cut_U2_mu4p6_phase_singlet}, in that they both break translational symmetry along the length of the cylinder with a doubling of the unit-cell. Due to the breaking of translational symmetry, this system is not ammendable to the kind of point-group symmetry analysis performed in Appendix II.

\begin{figure}[H]
\centering
 \subfigure{
    \includegraphics[width=0.75\linewidth]{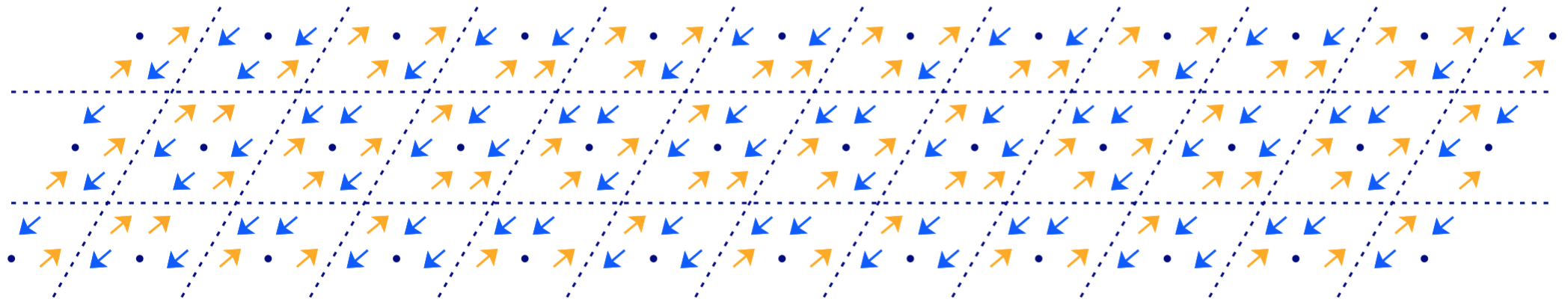}} 
\subfigure{
    \includegraphics[width=0.85\linewidth]{"phase_legend"}
    \label{fig:phase_legend}}
\caption{The triplet superconducting phase structure, Arg$(\langle c^{\dagger}_{i\uparrow}c^{\dagger}_{j\downarrow} + c^{\dagger}_{i\downarrow}c^{\dagger}_{j\uparrow} \rangle)$, for all nearest-neighbors with $U = +2$ for our $2\times 24$ lattice with a uniform edge-field, periodic boundary conditions along the short direction, and open boundary conditions along the long direction. For visibility, we truncate the plot so that only the half farthest from the edge field is shown. The arrow points in the direction of the phase and is also colored according to the argument for visibility. Note that this method of plotting possesses a redundency in that $\Delta^{triplet}_{ij} = - \Delta^{triplet}_{ji}$.}
\label{fig:L36cut_U2_mu4p6_phase_triplet}
\end{figure}

\clearpage

\section{Appendix V: Effects of Chemical Potential on the PDW}

We explore the role of the chemical potential in the PDW-like structure seen in the repulsive Hubbard regime. To this end we shift the chemical potential from  $\mu = 4.6$ to $\mu = 6.0$ for a fixed Hubbard interaction strength of $U = +2$. The resultant bond-centered superconducting pairing for the singlet channel is provided in Fig. A\ref{fig:L36cut_U2_mu6_phase_singlet}. Herein we see a similar behavior to the $\mu = 4.6$ case in that there is translational symmetry breaking along the length of the cylinder. However, rather than a doubling of the unit-cell, we see a tripling of the unit-cell. This tripling also occurs in the triplet channel which is not shown.
We note that the apparent randomness on the left side of the plot isn't due to edge field effects since this is the third of the lattice away from the probe and instead is an artefact of the small amplitude there relative to numerical convergence.

\begin{figure}[H]
\centering
\includegraphics[width=0.75\linewidth]{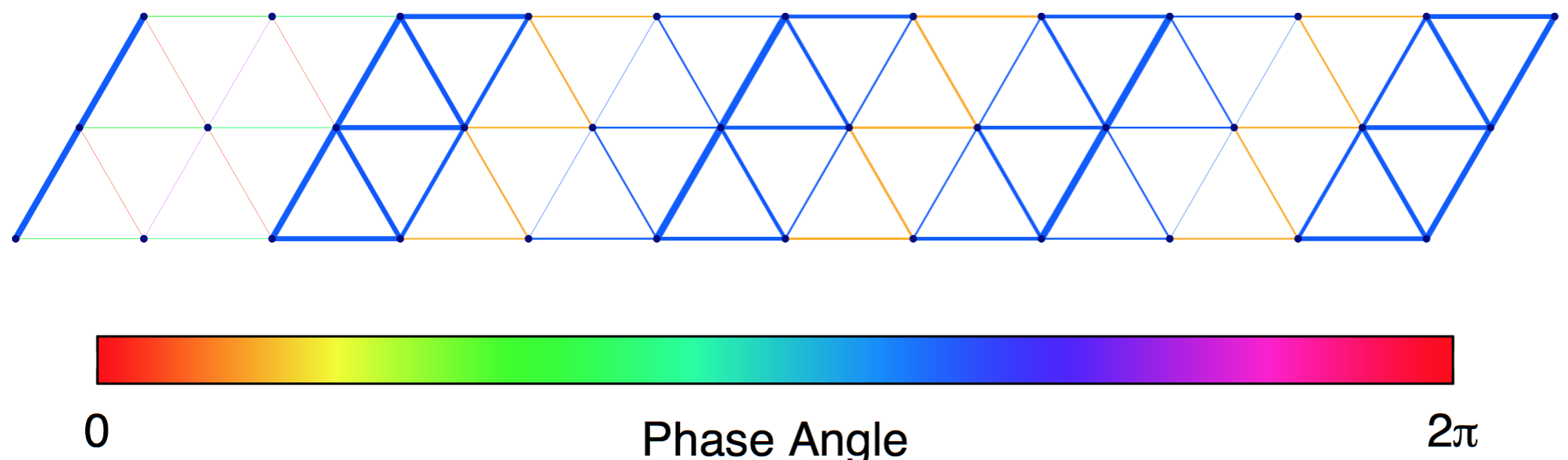}
\caption{The phase of the bond-centered singlet superconducting order parameter, $\Delta^{singlet}_{\langle i,j \rangle}$ for nearest-neighbors on the $3\times 36$ lattice with random pair-edge-field where now a larger chemical potential of $\mu = 6.0$ is used rather than that in the main text, $\mu = 4.6$. Here, the line thickness is proportional to the amplitude. The lattice has periodic boundary conditions along the short direction and open boundary conditions along the long direction.}
\label{fig:L36cut_U2_mu6_phase_singlet}
\end{figure}

\end{widetext}




\end{document}